\documentclass[pra,twocolumn,showpacs,amsmath,amssymb]{revtex4-1}
\usepackage{bm}
\usepackage[pdftex]{graphicx}

\begin{document}
\title{Phase separation of quantized vortices in two-component miscible Bose--Einstein condensates in a two-dimensional box potential}
\author{Junsik Han$^1$}
\author{Makoto Tsubota$^{1,2,3}$}
\affiliation{$^1$Department of Physics, Osaka City University, 3-3-138 Sugimoto, Sumiyoshi-ku, Osaka 558-8585, Japan}
\affiliation{$^2$The Advanced Research Institute for Natural Science and Technology, Osaka City University, 3-3-138 Sugimoto, Sumiyoshi-ku, Osaka 558-8585, Japan}
\affiliation{$^3$Nambu Yoichiro Institute of Theoretical and Experimental Physics (NITEP), Osaka City University, 3-3-138 Sugimoto, Sumiyoshi-ku, Osaka 558-8585, Japan}
\date{\today}


\begin{abstract}
The dynamics of quantized vortices in two-dimensional two-component miscible Bose--Einstein condensates (BECs) trapped by a box potential has been numerically studied using the Gross--Pitaevskii model.
We have discovered a novel phenomenon where the vortices of the two components spatially separate from each other, which we call the phase separation of the distribution of vortices.
This phase separation occurs when the inter-component coupling is strong.
Onsager vortices, on the other hand, are formed in both components when the inter-component coupling is weak.
We distinguish between Onsager vortices and phase-separated vortices by two types of effective distances between vortices.
The dependence of the transition between the Onsager vortices and phase-separated vortices on the inter-component interaction is also studied.
\end{abstract}

\maketitle


\section{Introduction}\label{sec:Introduction}

In turbulence, the velocity field of a fluid is chaotic spatially and temporally.
This turbulent velocity field of fluids comes from the complex dynamics of a multitude of vortices. 
This is unpredictable in the sense that the dynamics is very sensitive to the initial conditions, and we cannot understand the overall picture of turbulence by calculating the velocity field at a particular time and position.
Nevertheless, Taylor built a framework to study turbulence by introducing the idea that turbulence is homogeneous and isotropic on scales smaller than the system size \cite{Taylor35}.
Based on this idea, statistical laws have been studied.
Turbulence demonstrates universal statistical laws, which depend on the spatial dimension of the system \cite{Davidson15}.
Three-dimensional (3D) turbulence supports a direct cascade of energy transfers from large to small spatial scales.
The direct cascade is believed to be supported by the Richardson cascade, in which large vortices are broken into smaller ones.
Then, its energy spectrum obeys the Kolmogorov's -5/3 power law \cite{Kolmogorov41_1,Kolmogorov41_2}.
On the other hand, two-dimensional (2D) turbulence behaves in a stark contrast to 3D turbulence.
Onsager predicted the spontaneous formation of large-scale long-lived vortices, which are called Onsager vortices in two-dimensional (2D) turbulence \cite{Onsager49,Eyink06}.
Kraichnan predicted that 2D classical turbulence shows an inverse energy cascade from small to large spatial scales \cite{Kraichnan67}.

When we try to study the Richardson cascade in classical turbulence (CT), we face a great difficulty, because it is very difficult to distinctly define each vortex.
Quantum turbulence (QT) has the advantage that quantized vortices are a topological defect.
This means that quantized vortices are well-defined both in 3D and in 2D systems of QT \cite{Tsubota13,Tsatsos16}.
The atomic Bose--Einstein condensate (BEC) is a typical quantum liquid system and has the following advantages for studying QT.
First, the diluteness of the BEC gas makes the vortex core relatively large.
The size of the core becomes the coherence length $\xi$ of order $\mu {\rm m}$, so that optical techniques make the vortex core visible. 
Secondly, the BEC is well-controlled experimentally, for example, the intensity of the atom--atom interaction can be changed using the Feshbach resonance \cite{Inouye98}.
Finally, we can use the Gross-Pitaevskii (GP) equation, which is based on the mean field approximation, to treat the dynamics of the condensate \cite{PethickSmith08}.
The GP equation describes a lot of experimental results quantitatively \cite{Tsatsos16,Tsubota17}.

Three-dimensional QT also shows the direct energy cascade and the Kolmogorov's -5/3 power law, which can support the Richardson cascade process in the system \cite{Kobayashi05,Kadokura18}.
Then the vortices form a tangle and reconnect with each other.
A quantized vortex reconnection is a common phenomenon in which two vortex lines interact, antiparallel vortices approach to each other, connect at a given point, and exchange tails with the emission of phonons, which supports the direct energy cascade.

In addition, several experimental, numerical, and theoretical studies were conducted to determine whether Onsager vortices are formed in 2D QT \cite{Neely13,Seo17,Gauthier18,Johnstone18,Numasato10,White12,Reeves13,Billam14,Simula14,Billam15,Groszek16,Yu16}.
Specifically, Groszek {\it et al.} concluded that the formation of Onsager vortices depends on the trapping potential and the vortex-antivortex pair annihilation by the numerical simulations of the GP model.
Vortex-antivortex annihilation occurs only in 2D (or quasi-2D) systems and is an extreme case of vortex reconnection, which underpins the mechanism of evaporative heating of the vortex.

These studies addressed one-component BECs.
On the other hand, two-component BECs were also studied numerically in 2D \cite{Kasamatsu03,Kasamatsu05,Eto11,Nakamura12,Karl13_1,Kasamatsu16} and 3D \cite{Takeuchi10,Karl13_2}, and novel phenomena that are absent in one-component BECs were observed.
For example, vortices form interlocked triangular lattices, square lattices, or interwoven serpentine vortex sheets in 2D rotating condensates \cite{Kasamatsu03}. 
These phenomena originate from the competition between the intra-component and inter-component couplings of the condensates.
The intra-component coupling of the condensates is the interaction between atoms of the same condensate component, denoted by $g_{kk} \ (k = 1,2)$. 
The inter-component coupling of the condensates is the interaction between atoms of different condensate components, denoted by $g_{12}$.
Here, two components are miscible when $\sqrt{g_{11}g_{22}} > g_{12}$ and phase-separated when $\sqrt{g_{11}g_{22}} < g_{12}$.

These two types of coupling give rise to two types of vortex interaction: intra-component and inter-component interactions.
The intra-component interaction of the vortices is the interaction between vortices belonging to the same condensate component, whereas the inter-component interaction of the vortices means the interaction between vortices belonging to two different components.
The energy of the intra-component interaction of the vortices is expressed as 
\begin{equation}
\epsilon^{\rm intra}_{ij} = \frac{2\pi s_{i} s_{j} \hbar^{2} n}{m} \ln \frac{2R_{0}}{R_{ij}}, \label{intracomponent_interaction_energy}
\end{equation}
where $n$ is the density of the condensates, $m$ is the mass of the involved atom, $R_{ij}$ is the distance between the $i$-th and $j$-th vortices, and $R_{0}$ is the radius of the potential. \cite{PethickSmith08}.
The circulation signs of the $i$-th and $j$-th vortices are denoted by $s_{i}$ and $s_{j}$, respectively ($s_{i,j} = \pm 1$). 
If these circulation signs are equal, then the intra-component interaction is repulsive, and if the signs are different, it is attractive.
The energy of the inter-component interaction of the vortices is 
\begin{equation}
\epsilon^{\rm inter}_{ij} = \frac{\pi \hbar^{4} g_{12}}{4 m_{1} m_{2} (g_{11}g_{22} - g^{2}_{12})} \frac{\ln \frac{R'_{ij}}{\xi}}{R'^{2}_{ij}}, \label{intercomponent_interaction_energy}
\end{equation} 
where $m_{k}$ is the mass of the atom of the $k$-th component, $2R'_{ij}$ is the distance between $i$-th and $j$-th vortices belonging to two different components, and $\xi$ is the coherence length \cite{Eto11}.
Then, whether this interaction is repulsive or attractive, depends on signs of $g_{12}$ and $g_{11}g_{22} - g^{2}_{12}$.
The two types of interaction (Eqs. (\ref{intracomponent_interaction_energy}) and (\ref{intercomponent_interaction_energy})) show a different dependence on the distance between the vortices.
The presence of two types of interaction makes the vortex behavior in two-component BECs more complex and rich than in one-component BECs.

Two-component BECs were also studied experimentally.
In 2D system, the formation of vortex \cite{Matthews99}, the dynamics of a few vortices \cite{Seo15,Seo16}, the relaxation dynamics of turbulent condensates \cite{Seo16}, and interlaced square vortex lattices in a rotating system \cite{Schweikhard04} were studied.
However, there are no studies focused on the formation of Onsager vortices.

Here, we study the two-component BEC system.
We have done a preliminary study and then discovered a novel phenomenon, which we call the phase separation of the distribution of vortices \cite{Han18}.
The main motivation of this work is to elucidate the dependence of the transition between the formation of Onsager vortices and the phase separation of the distribution of vortices on $g_{12}$. 
We expect that the phase separation of the distribution of vortices is characteristic of the dynamics of multicomponent BECs.

In Sec. \ref{sec:Model}, we introduce the numerical model.
In Sec. \ref{sec:Results}, we present the results of our simulation.
We introduce the phase separation of the distribution of vortices in Sec. \ref{sec:phase_separation}.
In Sec. \ref{sec:distance}, we introduce two types of distances between vortices to distinguish the formation of Onsager vortices and phase separation of the distribution of vortices.
Then in Sec. \ref{sec:energy}, we discuss the dependence of the dynamics of vortices on the energies of intra-component and inter-component interactions of vortices.
They also depend on the inter-component coupling $g_{12}$ and the distances between the vortices.
Sec. \ref{sec:Conclusions} is devoted to discussion and conclusions.


\section{Model} \label{sec:Model}

In this study, we address 2D two-component BECs trapped by a box potential.
The $k$-th component of the BEC can be described by a macroscopic wave function $\psi_{k} = \sqrt{n_{k}({\bm r},t)}e^{\imath \phi_{k}({\bm r},t)}$, where $n_{k}({\bm r},t)$ is the density of the condensate, and $\phi_{k}({\bm r},t)$ is its phase.
These wave functions obey the GP equations
\begin{equation}
\begin{array}{l}
 \imath\hbar\frac{\partial}{\partial t}\psi_{k}({\bm r},t) \\
 \\
 = \left[\frac{-\hbar^{2}}{2m_{k}}\nabla^{2} + V_{\rm trap}({\bm r}) + \sum_{k'=1,2} g_{kk'}|\psi_{k'}({\bm r},t)|^{2}\right] \psi_{k}({\bm r},t). \\
 \\
   \hspace{1pc}(k = 1, 2)\label{eq:GPEs}
\end{array}
\end{equation} 
Here we choose $m_{1} = m_{2} = m$, $g_{11} = g_{22} = g$ and consider a circular box potential, which was also realized experimentally \cite{Gaunt13,Navon16},
\begin{equation}
V_{\rm trap}({\bm r}) = \left \{
\begin{array}{l}
V_{0} \hspace{1pc}(|{\bm r}| > R_{0}) \\
0 \hspace{1.5pc}(|{\bm r}| < R_{0})
\end{array} 
\right.,\label{eq:box_potential}
\end{equation}
with potential radius $R_{0}$ and potential height $V_{0}$.

To create the initial state, we imprint vortices in the condensate by multiplying the wave function by a phase factor $\Pi^{N_{v}}_{i}\exp(\imath \phi_{i})$, with $\phi_{i}(x,y) = s_{i}\arctan[(y-y_{i})/(x-x_{i})]$. 
Here, the coordinates $(x_{i},y_{i})$ refer to the position of the $i$-th vortex, and they are chosen randomly.
We take the number of initial vortices $N_{v} = 120$.
After imprinting, the wave function evolves in imaginary time to establish the structure of the vortex cores.
After the vortex cores appear clearly, we treat the state as an initial state.
Subsequently, we solve Eq. (\ref{eq:GPEs}) in real time using the Fourier and Runge--Kutta methods on a $512 \times 512$ spatial grid.

Vortices are identified by finding the phase singularities of the wave function. 
The number of vortices is counted in the $|{\bm r}| < 0.9 R_{0}$ region to avoid counting ghost vortices in the low-density region \cite{Tsubota02}.
In order to characterize the formation of the Onsager vortices, we also calculate the amplitude of the dipole moment of the vortex distribution, defined as $d=|{\bm d}|=|\sum_{i}q_{i}{\bm r}_{i}|$ \cite{Groszek16}.
Here ${\bm r}_{i}$ is the position of the $i$-th vortex, $q_{i} = s_{i}\kappa = s_{i}h/m$ is its charge with $s_{i}=\pm1$.
If the vortices are distributed uniformly, $d$ seldom grows. 
If like-sign vortices form an Onsager vortex, $d$ develops into a finite value.


\section{Results} \label{sec:Results}

\subsection{Onsager vortices and phase separation of the distribution of vortices}\label{sec:phase_separation}

\begin{center}
\begin{figure}[h]
\includegraphics[width=20pc, height=10pc]{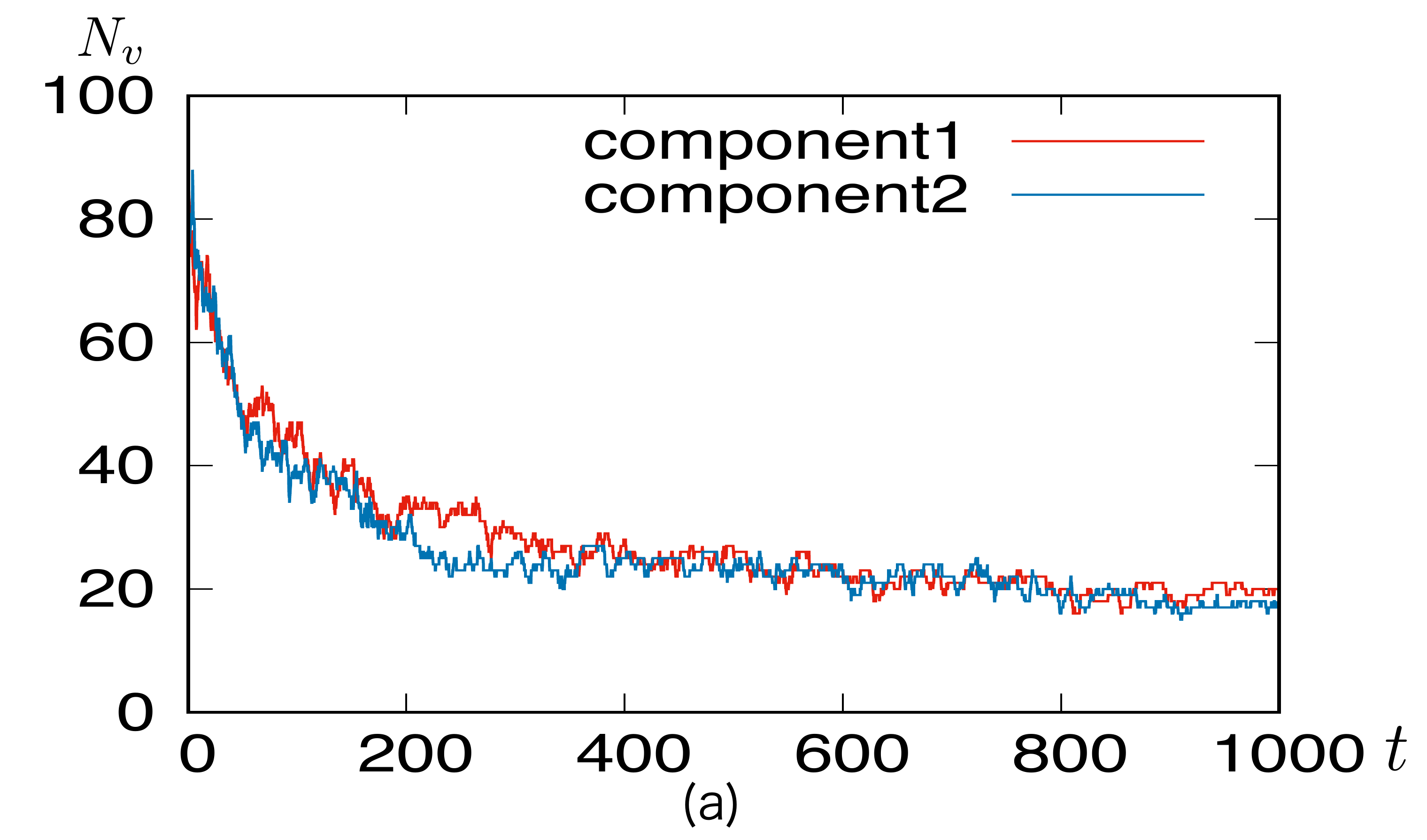}
\includegraphics[width=20pc, height=10pc]{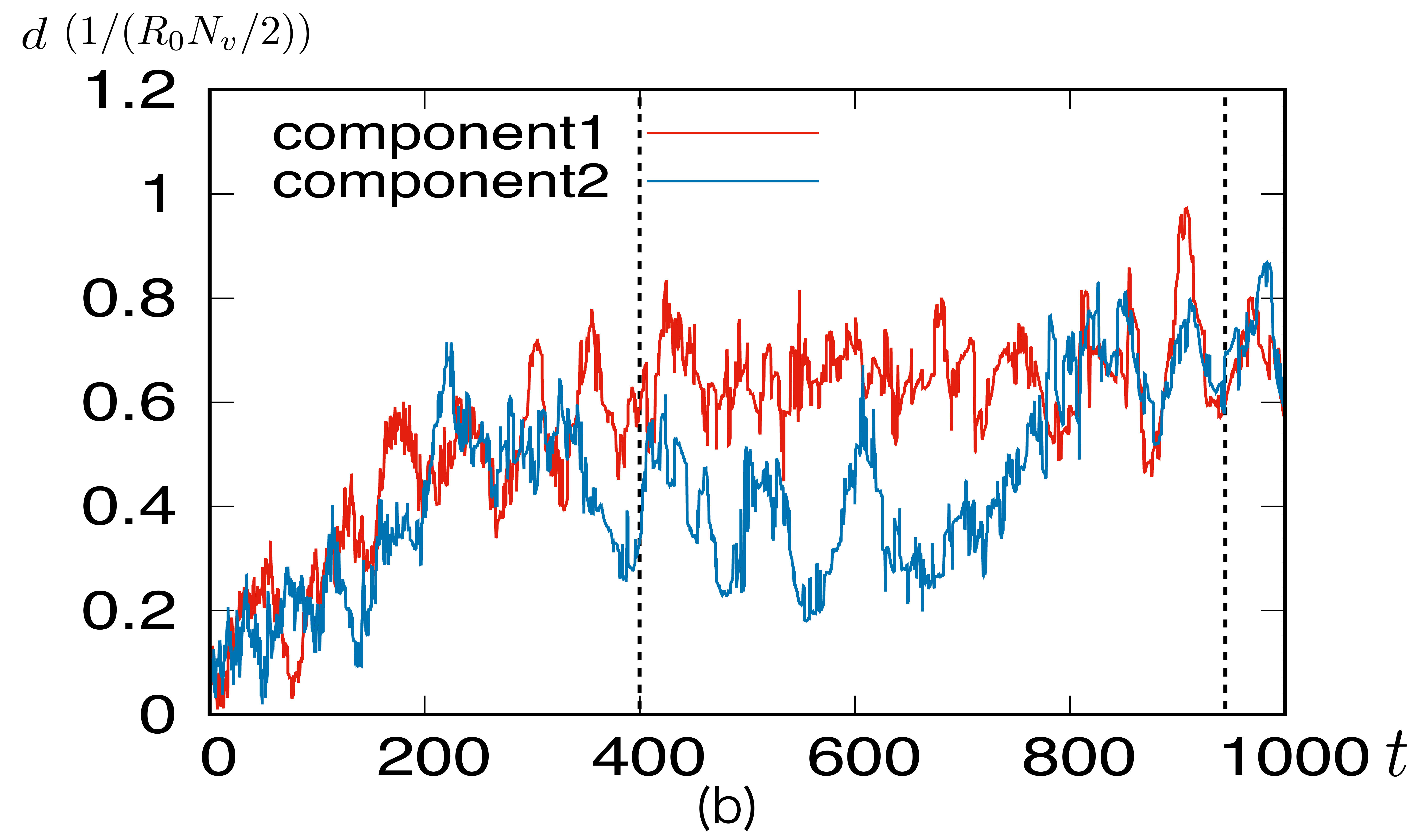}
\caption{\label{fig:2C_Nv_dipole_101} The panels show the time evolution of (a) the number of vortices $N_{v}$, and (b) the dipole moment $d$ in a two-component BEC.
The inter-component coupling $g_{12} = 0.1g$ is constant.
The dipole moment is normalized by the radius of the potential ($R_{0}$), the vortex charge ($\kappa$), and the number of vortex pairs ($N_{v}/2$).
The two vertical dotted lines correspond to the time frames displayed in Fig. \ref{fig:2C_point_101}\\
{\it Source}: Paper of Han \cite{Han18}.}
\end{figure}
\end{center}

\begin{center}
\begin{figure}[h]
\includegraphics[width=20pc, height=11.25pc]{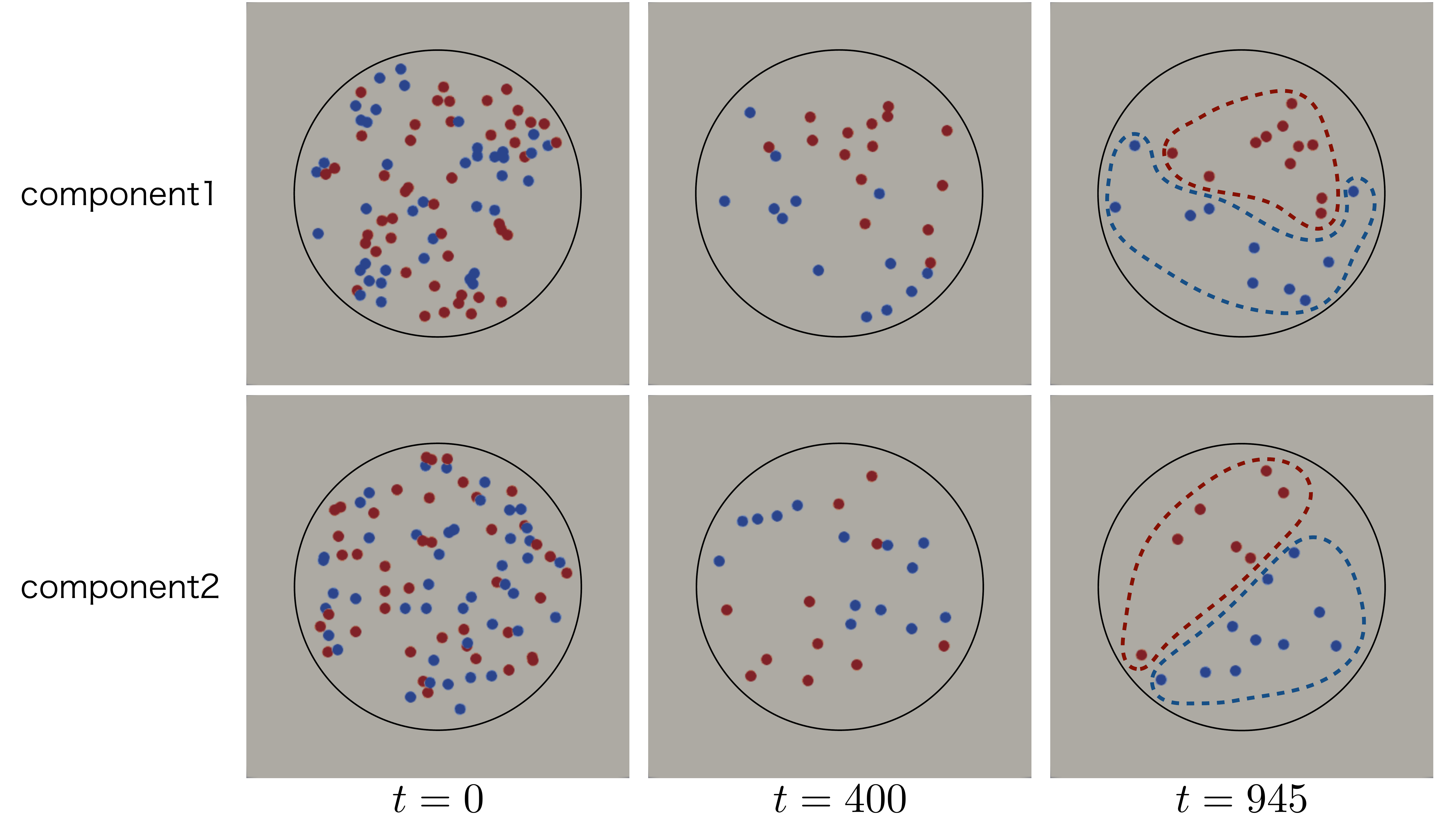}
\caption{\label{fig:2C_point_101} Vortex distribution in a two-component BEC at $t = 0$, $400$, and $945$.
The inter-component coupling $g_{12} = 0.1g$ is constant.
Red and blue points are vortices and antivortices, and black circles represent the boundary of the condensate. 
Two Onsager vortices appear at $t = 945$ in both components; one consists of vortices (surrounded by the red dotted line), and the other consists of antivortices (surrounded by the blue dotted line).}
\end{figure}
\end{center}

\begin{center}
\begin{figure}[h]
\includegraphics[width=20pc, height=10pc]{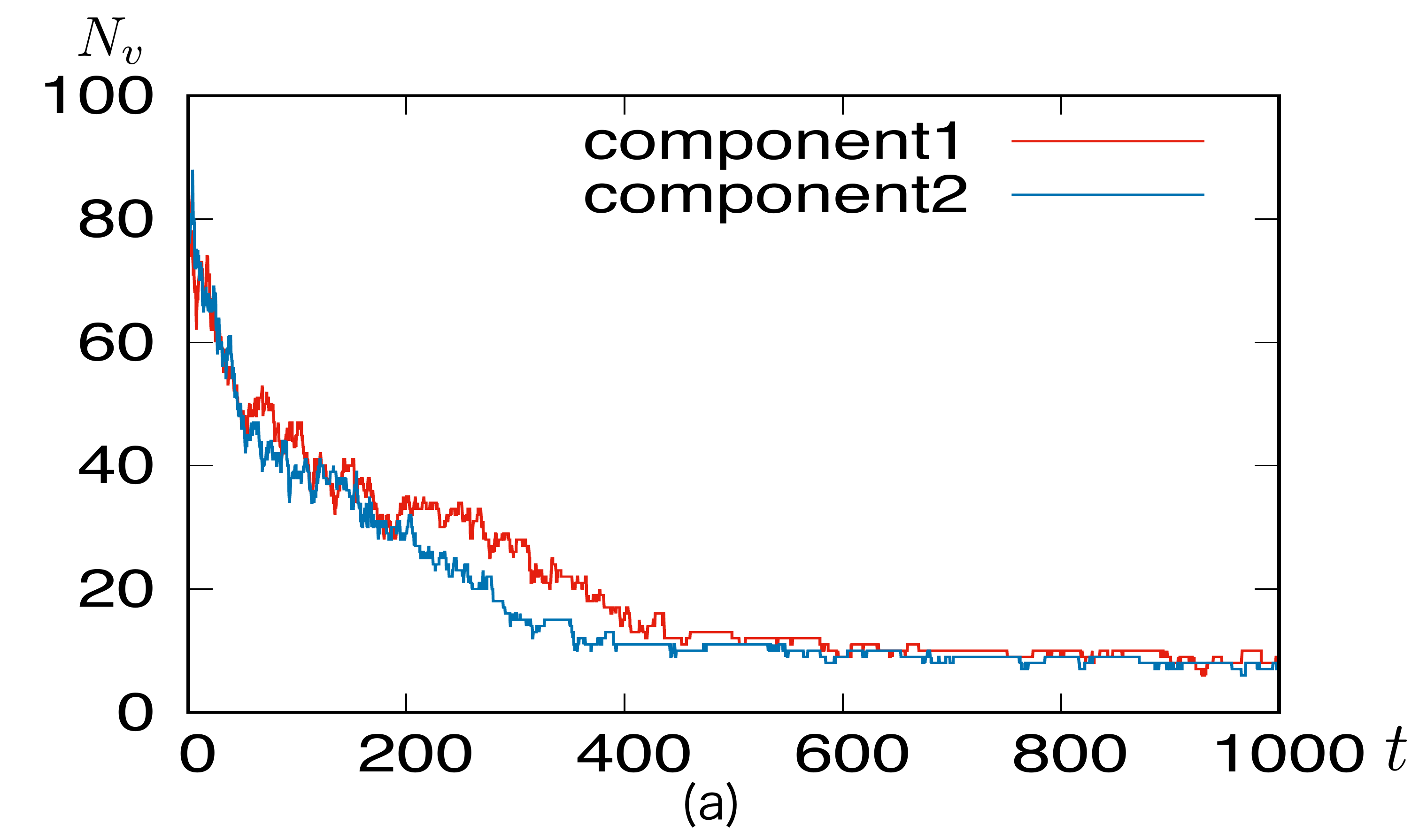}
\includegraphics[width=20pc, height=10pc]{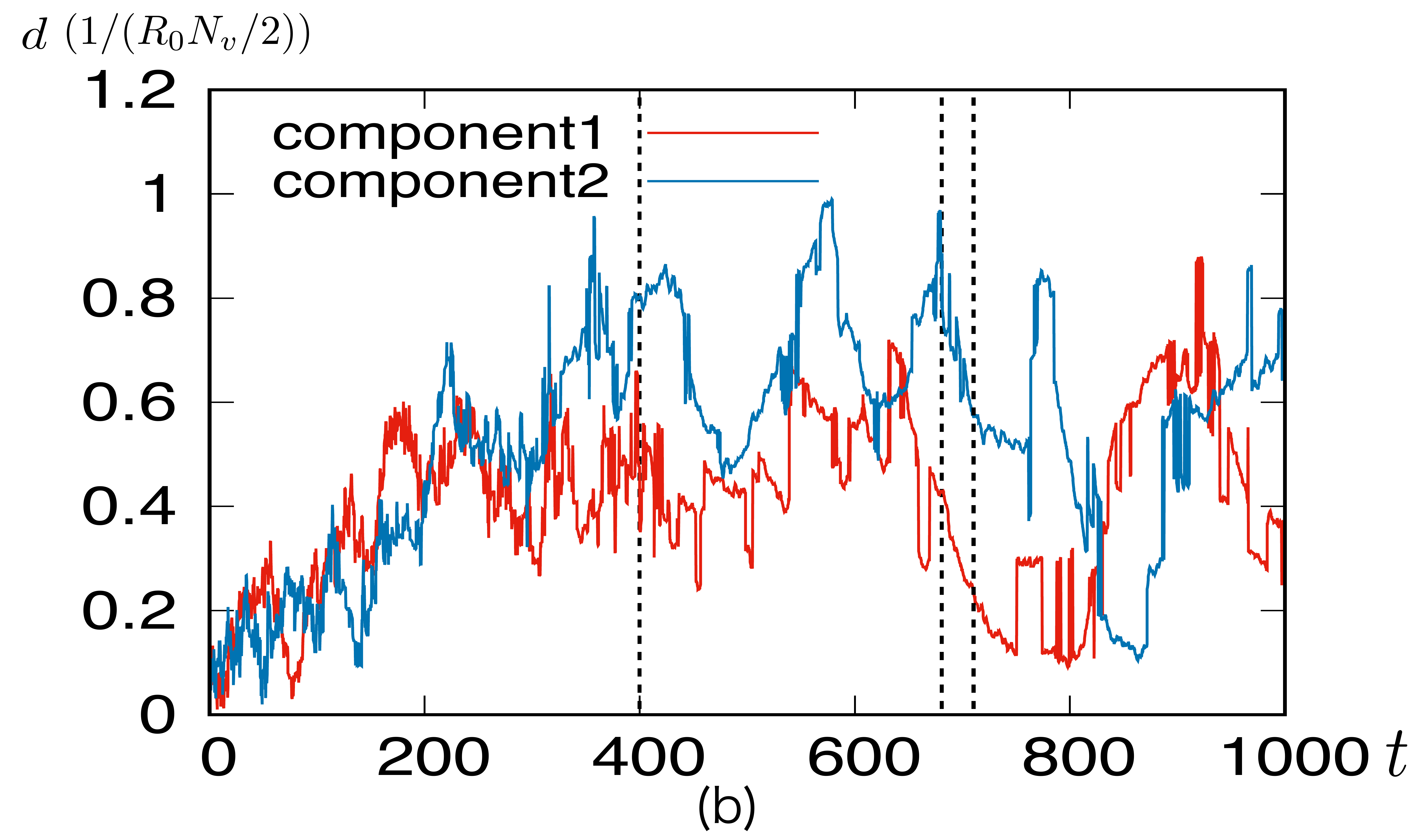}
\caption{\label{fig:2C_Nv_dipole_107} The panels show the time evolution of (a) the number of vortices $N_{v}$ and (b) the dipole moment $d$ in two-component BECs.
The inter-component coupling $g_{12}$ changes from $0.1g$ to $g_{12} = 0.7g$ at $t = 250$.
The dipole moment is normalized by the radius of the potential ($R_{0}$), the vortex charge ($\kappa$), and the number of vortex pairs ($N_{v}/2$).
The two vertical dotted lines correspond to the time frames displayed in Fig. \ref{fig:2C_point_107}. \\
{\it Source}: Paper of Han \cite{Han18}.} \end{figure}
\end{center}

\begin{center}
\begin{figure}[h]
\includegraphics[width=20pc, height=11.25pc]{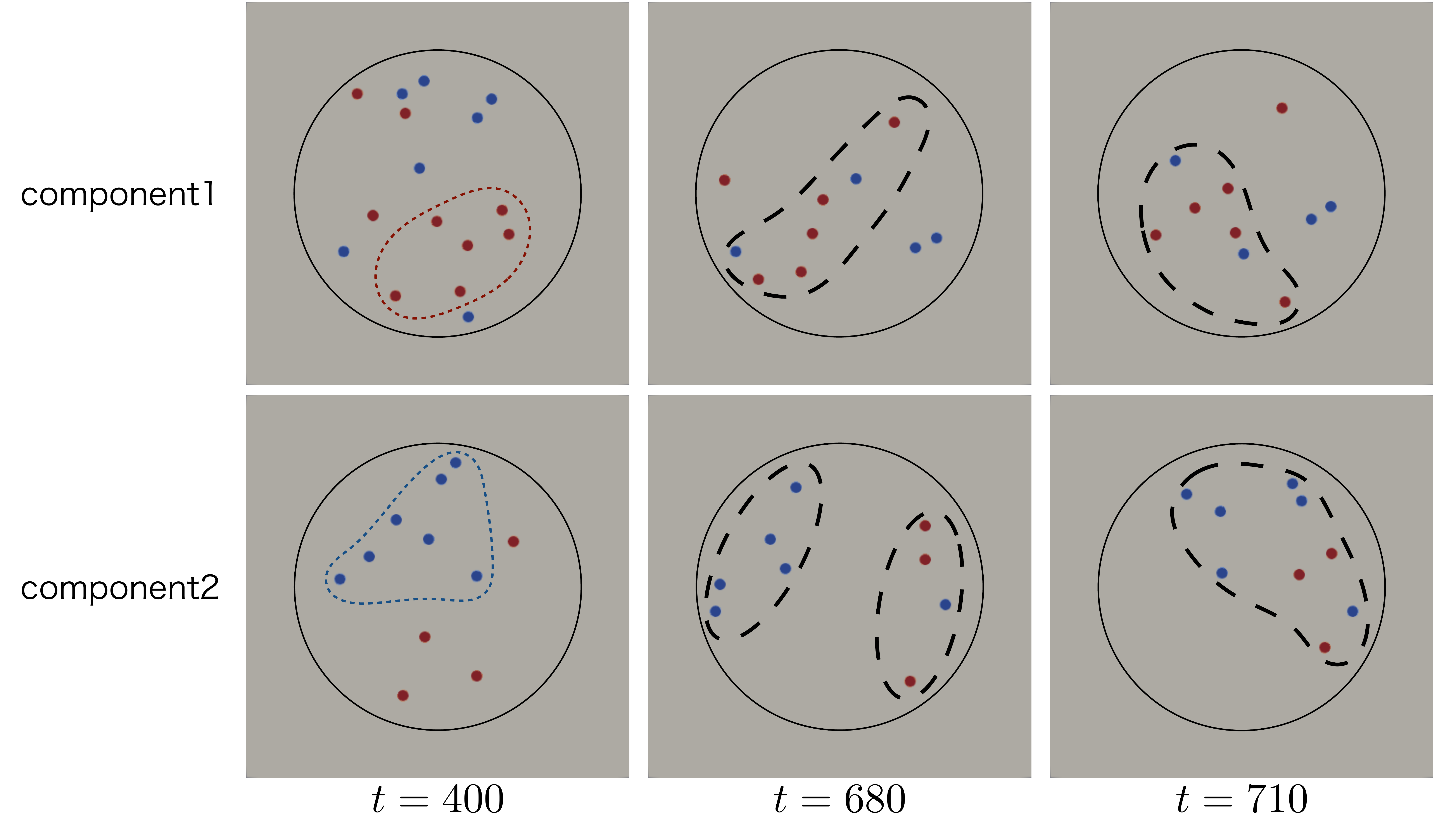}
\caption{\label{fig:2C_point_107} Vortex distribution in a two-component BEC at $t = 400$, $680$, and $710$.
The inter-component coupling $g_{12}$ changed from $0.1g$ to $g_{12} = 0.7g$ at $t = 250$.
Red and blue points are vortices and antivortices, and black circles represent the boundary of the condensate. 
The tendency of clustering of the like-sign vortices is confirmed at $t = 400$ (surrounded by red and blue dotted lines).
Dotted red and blue lines surround clusters of vortices and antivortices, respectively.
At $t = 680$ and $t =710$, the cluster of vortices is in the region surrounded by the dashed black line for each component.
The phase separation of the distribution of vortices at $t = 710$ is the phase separation of Type ${\rm I}$, and that at $t = 680$ is the phase separation of Type ${\rm I\hspace{-.1em}I}$.}
\end{figure}
\end{center}

Here we initially choose $g_{12} = 0.1 g > 0$.
To focus on the dependence of the Onsager vortex formation on $g_{12}$, we consider two conditions: (i) $g_{12} = 0.1 g$ is constant, and (ii) we change $g_{12}$ from $0.1 g$ to $0.7 g$ at $t = 250$.
Under both conditions (i) and (ii), these two components are miscible.
Figures \ref{fig:2C_Nv_dipole_101} and \ref{fig:2C_Nv_dipole_107} show the time evolution of the number of vortices $N_{v}$ and dipole moment $d$ with conditions (i) and (ii).
The configuration of the vortices at the time points indicated in Figs. \ref{fig:2C_Nv_dipole_101} and \ref{fig:2C_Nv_dipole_107} with vertical dotted lines are shown in Figs. \ref{fig:2C_point_101} and \ref{fig:2C_point_107}, respectively.

We first focus on the result for condition (i).
Initially, the vortices of both components are randomly distributed (Fig. \ref{fig:2C_point_101} left column), and the dipole moment $d$ is almost $0$ (Fig. \ref{fig:2C_Nv_dipole_101}).
As time goes by, the $d$ of both components increases to a finite value with the decay of the number $N_{v}$ of vortices due to the vortex-antivortex annihilation.
Through this process, clusters of like-sign vortices develop, and, finally, two Onsager vortices appear at $t = 945$ in both components; one consists of vortices ($s_{i} = 1$), and the other consists of antivortices ($s_{i} = -1$) (Fig. \ref{fig:2C_point_101}, center and right column).
These results for each component are consistent with the previous simulation with one-component BECs \cite{Groszek16}.

Under condition (ii), we can also confirm the tendency of the formation of like-sign vortices in each component at $t = 400$, as in the case of condition (i) (Fig. \ref{fig:2C_point_107}, left column, dotted red and blue lines).
However, a novel phenomenon appears after sufficient time.
After $t \simeq 400$ (Fig. \ref{fig:2C_Nv_dipole_107}, left vertical dotted line), the development time of $d$ changes qualitatively.
This time dependence of $d$ after $t \simeq 400$ is due to the phase separation of the distribution of vortices.
This phase separation means that the vortices occupy some region smaller than the whole condensations; this region is called the vortex region (VR). 
Around $t = 680$ and $t = 710$, the typical influence of the phase separation of the distribution of vortices on $d$ can be confirmed (Fig. \ref{fig:2C_Nv_dipole_107}, center and right vertical dotted lines).
First, we focus on the distribution of VRs around $t = 710$.
The dipole moments $d$ of both components decrease considerably.
At this time, the vortices cluster in the region surrounded by the dashed black line for each component (Fig. \ref{fig:2C_point_107}, right column), which means that the VR in each component only becomes smaller.
We call this the phase separation of Type ${\rm I}$. 
This makes a significant decrease of $d$ of both component.

At approximately $t = 680$, $d$ of component 1 decreases, while $d$ of component 2increases.
At $t = 680$, the vortices cluster in the region surrounded by the dashed black line for each component, like at $t = 710$. 
At this time, the VR of component 2 is divided into two small VRs by the VR of component 1 (Fig. \ref{fig:2C_point_107}, center column).
Then the $d$ of component 1 decreases, while the $d$ of component 2 may increase.
This phase separation of the distribution of vortices accompanies the VR separation, which is called the phase separation of Type ${\rm I\hspace{-.1em}I}$.
Approximately at $t \simeq 570$, and $780$, we also confirm the phase separation of Type ${\rm I\hspace{-.1em}I}$.
This means that the transition between the phase separation of Type ${\rm I}$ and Type ${\rm I\hspace{-.1em}I}$ occurs in this system, and we cannot confirm the formation of Onsager vortices as under condition (i), even after sufficient time.

The difference in the dynamics of the vortices between the two conditions is due to the difference in $g_{12}$.
If $g_{12} = 0$, each component is independent and vortices belonging to one component are not affected by vortices belonging to the other component.
Hence, their dynamics are perfectly equivalent to the dynamics of a one-component BEC, when the initial distributions are the same.
If $g_{12}$ is small, as under condition (i), the weak inter-component interaction works between the vortices belonging to other components.
This weak inter-component interaction does not affect the dynamics of vortices so much, not preventing each component from forming Onsager vortices. 
On the other hand, a large $g_{12}$ like condition (ii) makes a strong inter-component interaction.
This strongly affects the dynamics of vortices and leads to the phase separation of the distribution of vortices.
In the following subsections, we indicate the dependence of these phenomena on $g_{12}$.


\subsection{Effective distance between the centers of gravity of the vortices}\label{sec:distance}

We attribute the transition from the formation of Onsager vortices to the phase separation of the distribution of vortices to the interaction between the vortices.
In order to examine their dependence on $g_{12}$, we also calculate the dynamics of the vortices with the condition that we change $g_{12}$ from $0.1g$ to $0, \ 0.3g, \ 0.5g, \ 0.6g$ at $t = 250$.

\begin{center}
\begin{figure}[h]
\includegraphics[width=20pc, height=10pc]{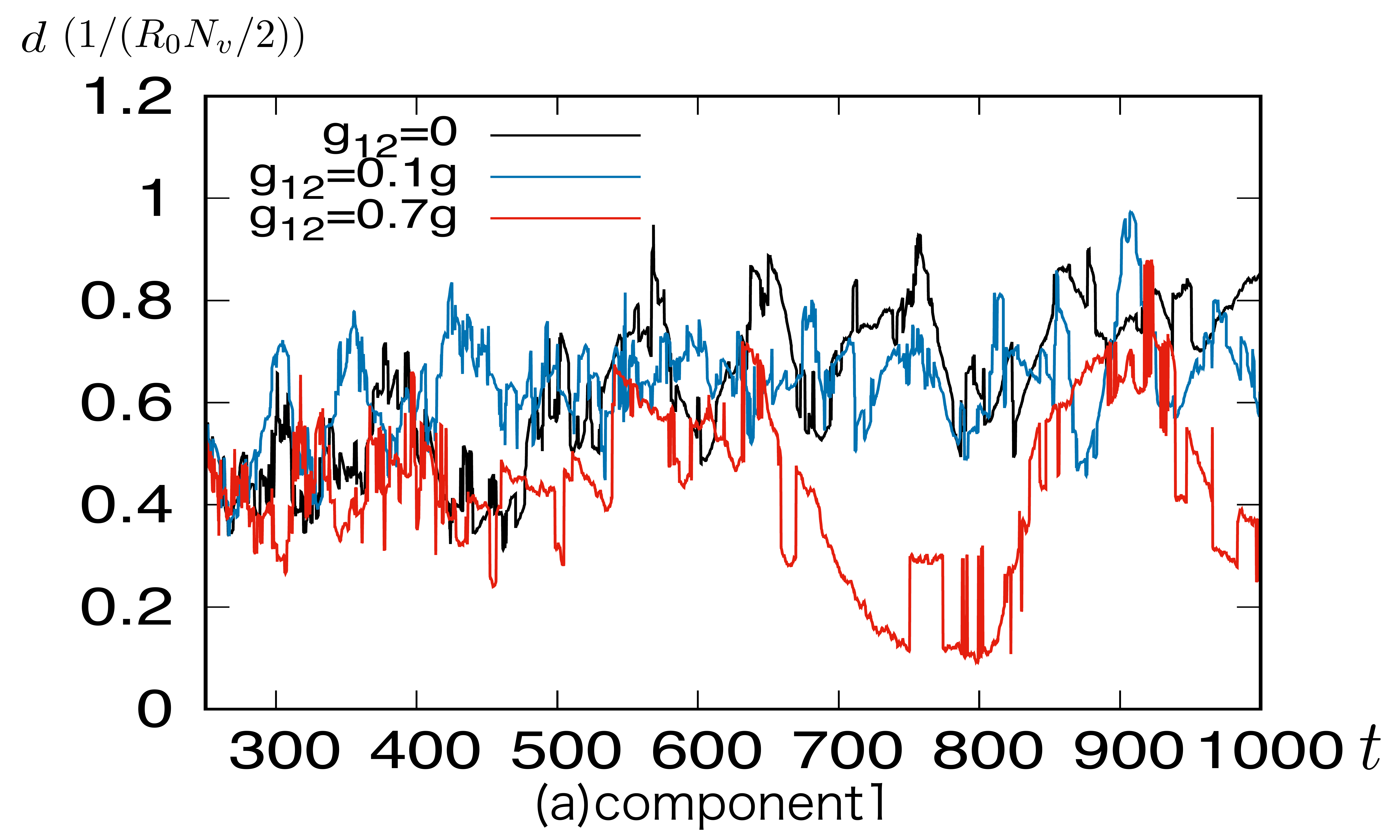}
\includegraphics[width=20pc, height=10pc]{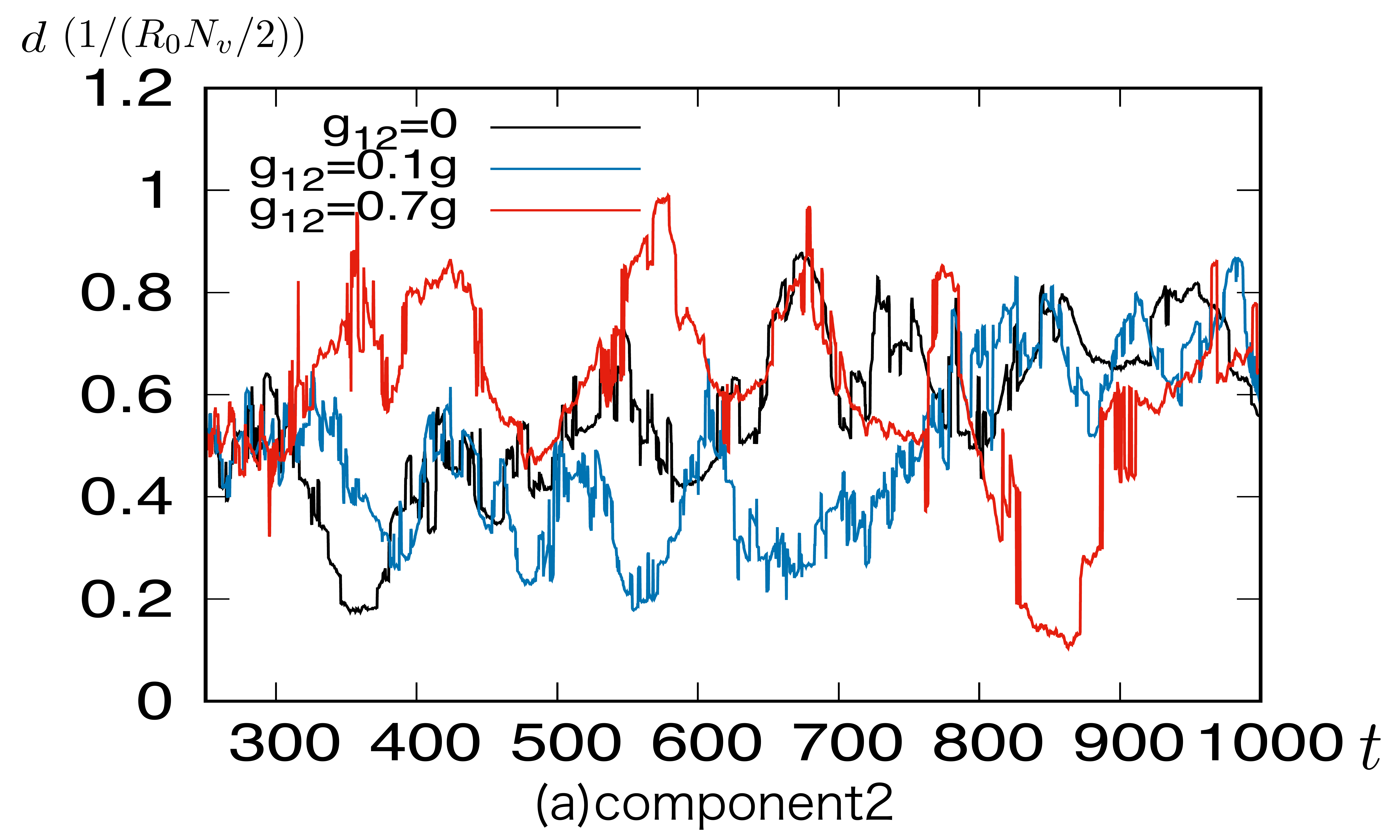}
\caption{\label{fig:2C_dipole_017} The panels show the time evolution of the dipole moment in (a) component1 and (b) component2.
The inter-component coupling values are $g_{12} = 0$ (black), $g_{12} = 0.1g$ (red), and $g_{12} = 0.7g$ (blue).
The dipole moment is normalized by the radius of the potential ($R_{0}$), the vortex charge ($\kappa$), and the number of vortex pairs ($N_{v}/2$).}
\end{figure}
\end{center}

Figure \ref{fig:2C_dipole_017} shows the time evolution of the dipole moment with $g_{12} = 0$, $0.1g$, and $0.7g$.
When $g_{12} = 0$, the states of each component are completely equivalent to the states of one-component BECs, and Onsager vortices are formed.
When $g_{12} = 0.1g$, Onsager vortices are formed in both components (Fig. \ref{fig:2C_point_101}, right column), and the development of $d$ is qualitatively similar to that at $g_{12} = 0$.
When $g_{12} = 0.7g$, the distribution of vortices is phase-separated (Fig. \ref{fig:2C_point_107}, center and right column), and the development of $d$ is qualitatively different from that at $g_{12} = 0$.
Then we can estimate the dynamics of the vortices by the difference between each $d_{l}$, where $d_{l} \ (l = 0, \ 0.1, \ 0.3, \ 0.5, \ 0.6, \ 0.7)$ means the dipole moment with $g_{12} = 0, \ 0.1g, \ 0.3g, \ 0.5g, \ 0.6g, \ 0.7g$.
We calculate the deviation of $d_{l}$ from $d_{0}$, defined as 
\begin{equation}
d^{\rm dev}_{l} = \frac{d^{\rm dev}_{l,1} + d^{\rm dev}_{l,2}}{2}, \label{dipole_deviation}
\end{equation} 
\begin{equation}
d^{\rm dev}_{l,k} = \frac{1}{750}\int^{1000}_{t = 250}(d(t)_{0,k} - d(t)_{l,k})^{2}dt, \ (k = 1 , 2) \label{dipole_deviation_component}
\end{equation} 
where $d(t)_{k}$ is the dipole moment of the $k$-th component at $t$.
If $d^{\rm dev}_{l}$ is small, the development of $d_{l}$ is similar to that of $d_{0}$, and this may be an indicator of the formation of Onsager vortices.
If $d^{\rm dev}_{l}$ is large, the development of $d_{l}$ differs from that of $d_{0}$, and this may be an indicator of the phase separation of the distribution of vortices.

\begin{center}
\begin{figure}[h]
\includegraphics[width=20pc, height=10pc]{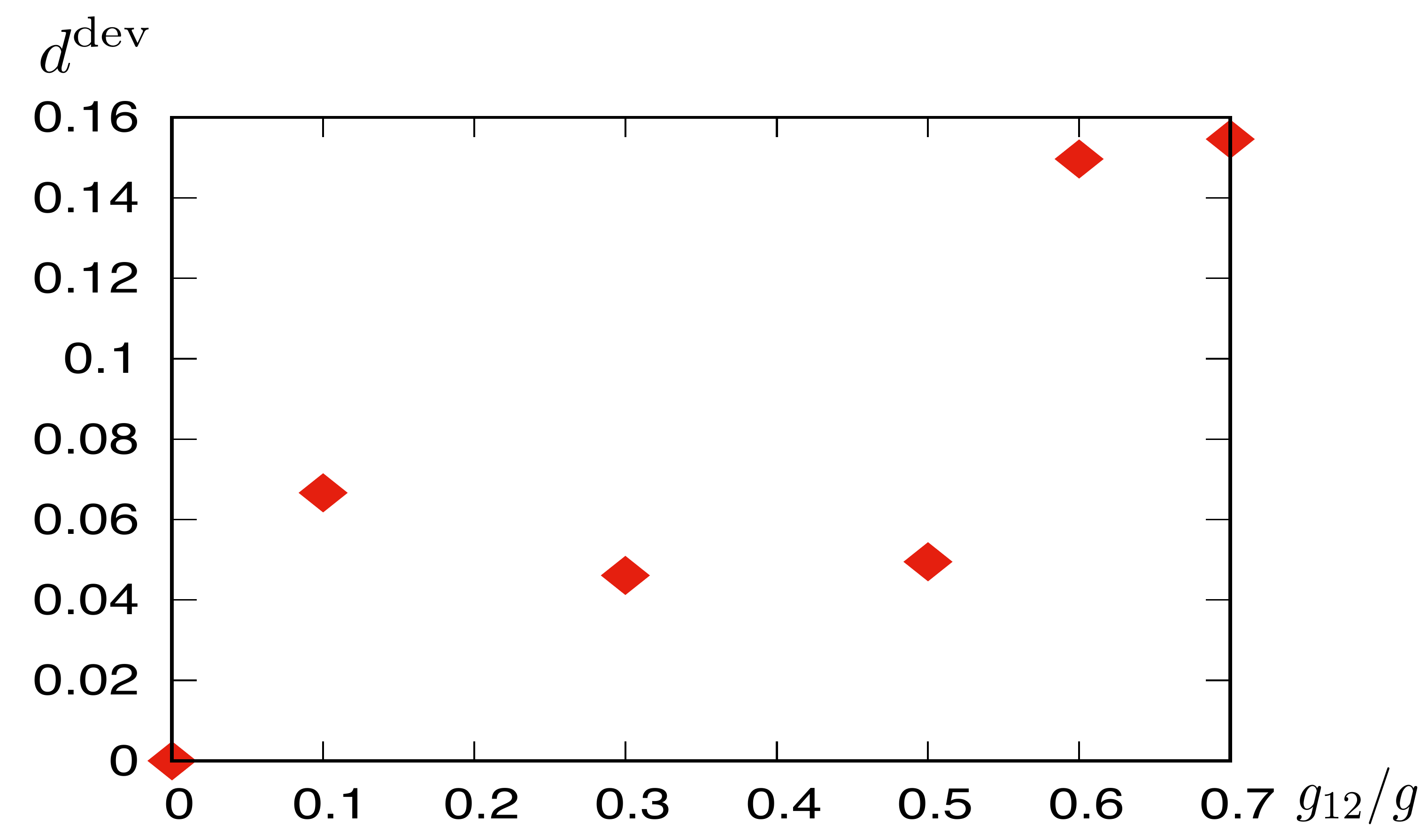}
\caption{\label{fig:2C_dipole_dev}The value of the deviation of dipole moment is $d^{\rm dev}_{l}$ for each $l$.
When $g_{12} \leqq 0.5g$, $d^{\rm dev}_{l}$ is small, and when $g_{12} \geqq 0.6g$, $d^{\rm dev}_{l}$ is large.}
\end{figure}
\end{center}

Figure \ref{fig:2C_dipole_dev} shows the value of $d^{\rm dev}_{l}$ for each $l$.
When $g_{12} \leqq 0.5g$, $d^{\rm dev}_{l}$ is small, but when $g_{12} \geqq 0.6g$, $d^{\rm dev}_{l}$ suddenly increases.
This means that the dynamics of vortices change drastically near $g_{12} \simeq 0.6$.
Onsager vortices appear at $g_{12} < 0.5g$, and the distribution of vortices is phase-separated at $g_{12} > 0.6$.

To distinguish the formation of the Onsager vortices and the phase separation of the distribution of vortices more clearly, we introduce two types of distance.
One of these is the average of distances between the center of gravity of the vortices (CGV) and all the vortices in one component, which means the effective size of VR and is denoted as $\tilde{L}_{\rm VR}$.
By introducing the CGV of the $k$-th component,
\begin{equation}
\tilde{\bm r}_{{\rm cg},k} = \frac{1}{N_{v}}\sum_{i}{\bm r_{i, k}}, \ (k=1,2) \label{cg}
\end{equation} 
we have
\begin{equation}
\tilde{L}_{{\rm VR},k} = \frac{1}{N_{v,k}}\sum_{i}\left| \tilde{\bm r}_{{\rm cg},k} - \bm r_{i,k} \right|, \label{Lk_VR}
\end{equation}
and the average of the two components is
\begin{equation}
\tilde{L}_{\rm VR} = \frac{\tilde{L}_{{\rm VR},1} + \tilde{L}_{{\rm VR},2}}{2}. \label{effective_L_mean}
\end{equation} 
The other is the distance between the CGVs of each component, which is defined by
\begin{equation}
\tilde{R}_{\rm cg} = \left| \tilde{\bm r}_{{\rm cg},1} - \tilde{\bm r}_{{\rm cg},2} \right|. \label{R_cg}
\end{equation}  
They represent a feature of the distribution of vortices.

\begin{center}
\begin{figure}[h]
\includegraphics[width=20pc, height=15.6pc]{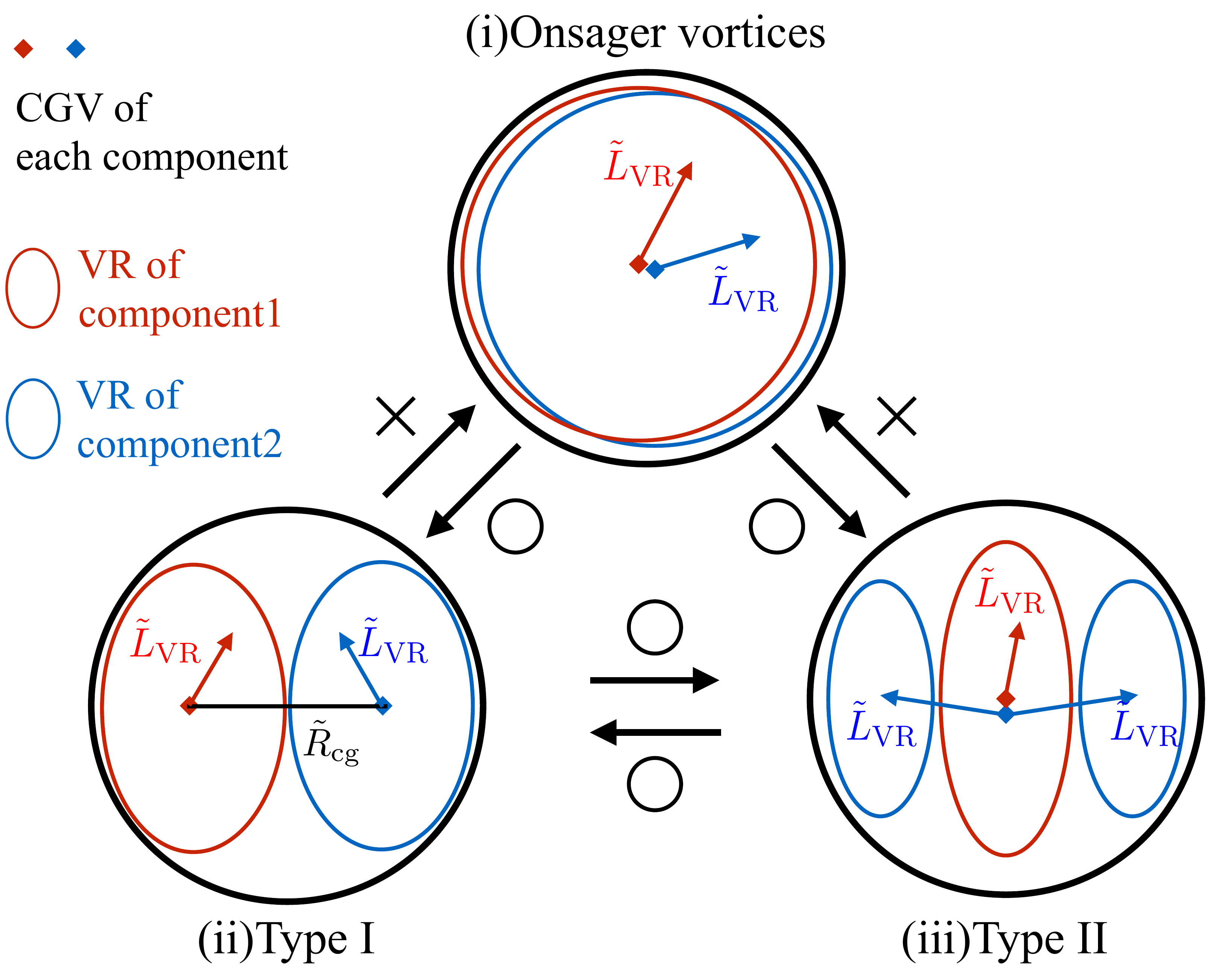}
\caption{\label{fig:R_cgp} Three schematic patterns of the VR in the following cases: (i) when the Onsager vortices are formed, (ii) when the vortices show the phase separation of Type ${\rm I}$ and (iii) Type ${\rm I\hspace{-.1em}I}$.
The black circle represents the boundary of the condensation, and the regions surrounded by red and blue lines represent the VR of component 1 and component 2.
The red and blue arrows denote the $\tilde{L}_{\rm VR}$ of each component, and the black line means $\tilde{R}_{\rm cg}$.
The distance $\tilde{R}_{\rm cg}$ takes some finite value in the Type $\rm I$, but it is too small in the Onsager vortices and Type ${\rm I\hspace{-.1em}I}$ to be shown here.
The arrows between the two patterns indicate the possibility of transition.
Circles and crosses next to them mean whether a transition can occur or not.}
\end{figure}
\end{center}

Figure \ref{fig:R_cgp} shows a cartoon of the VR when the Onsager vortices are formed or vortices show phase separation of the Type ${\rm I}$ and the Type ${\rm I\hspace{-.1em}I}$.
Here, $\tilde{L}_{\rm VR}$s of each component are illustrated by red and blue arrows.
The red and blue diamonds refer to the CGVs of the two components, and their distance is $\tilde{R}_{\rm cg}$.
This $\tilde{R}_{\rm cg}$ becomes very small in Onsager vortices or Type ${\rm I\hspace{-.1em}I}$ (Fig. \ref{fig:R_cgp}, top and bottom right), while it becomes large in Type ${\rm I}$ and is shown by the black line (Fig. \ref{fig:R_cgp}, bottom left).
Thus, phase separation of Type ${\rm I}$ can be distinguished from the formation of Onsager vortices and phase separation of Type ${\rm I\hspace{-.1em}I}$ by calculating the ratio $P_{\rm cg}$ of two types of distances, defined as
\begin{equation}
P_{\rm cg} = \frac{\tilde{R}_{\rm cg}}{\tilde{L}_{\rm VR}}. \label{P_cg}
\end{equation}
When $P_{\rm cg} > 1$, we can judge that the system shows the phase separation of Type ${\rm I}$.
When $P_{\rm cg} < 1$, the system should form Onsager vortices or cause phase separation of Type ${\rm I\hspace{-.1em}I}$.
From the results of numerical simulations, the transition between these three distributions of vortices are not necessarily reversible.
According to the numerical results, the transition from the Onsager vortices to the Type ${\rm I}$ or Type ${\rm I\hspace{-.1em}I}$ can occur easily, but the inverse process hardly occurs (Fig. \ref{fig:R_cgp} arrows).
The transition between the Type ${\rm I}$ and the Type ${\rm I\hspace{-.1em}I}$ occurs frequently.

\begin{center}
\begin{figure}[h]
\includegraphics[width=20pc, height=10pc]{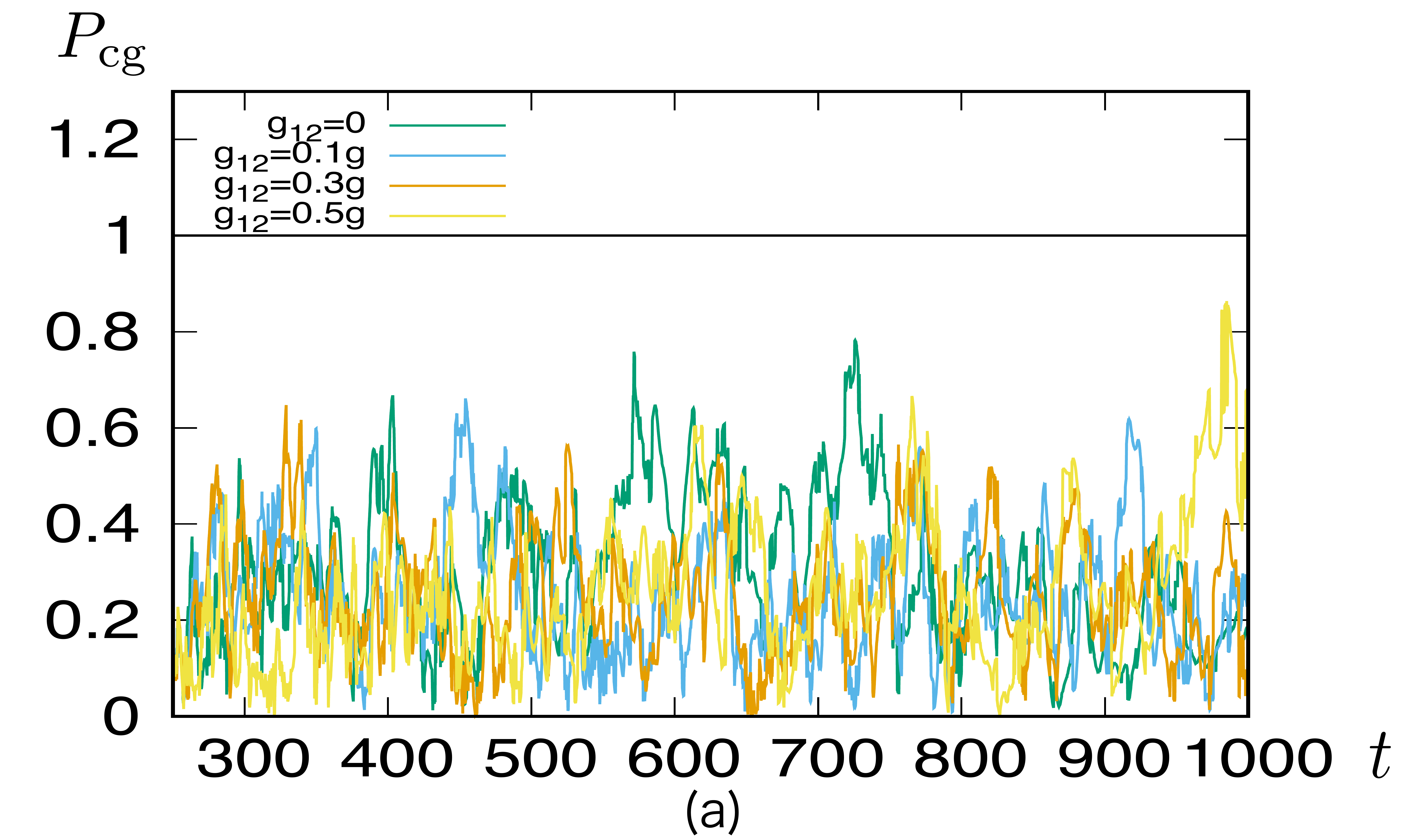}
\includegraphics[width=20pc, height=10pc]{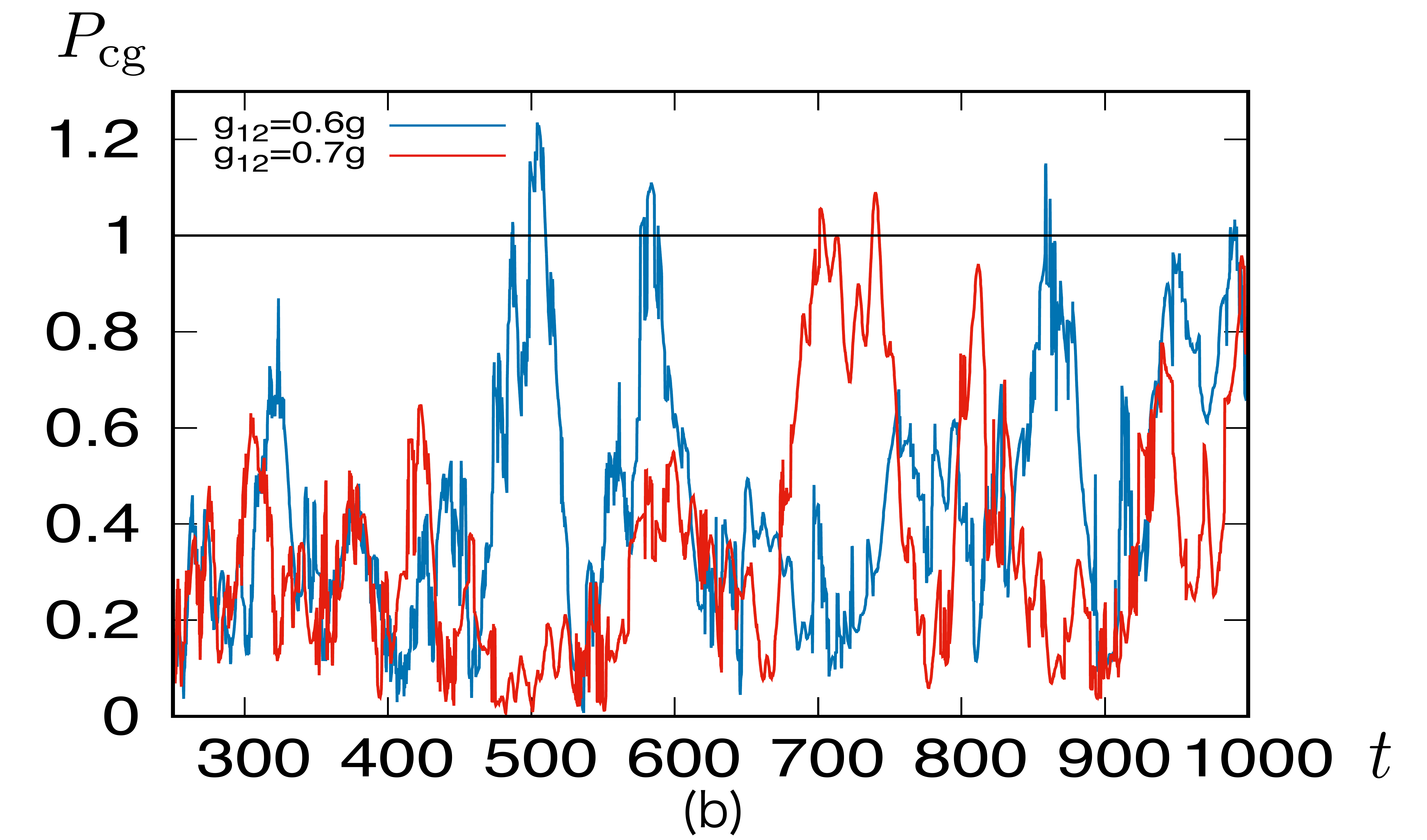}
\caption{\label{fig:cgp_rate} The time development of $P_{\rm cg}$.
Panel (a) shows the result when $g_{12} \leqq 0.5g$, and panel (b) shows the result when $g_{12} \geqq 0.6g$.}
\end{figure}
\end{center}

We can distinguish Onsager vortices and phase-separated vortices by the time dependence of $P_{\rm cg}$.
If $P_{\rm cg}$ sometimes exceeds unity, the system simply makes a transition between Type ${\rm I}$ and Type ${\rm I\hspace{-.1em}I}$ because the Onsager vortices do not appear when Type ${\rm I}$ or Type ${\rm I\hspace{-.1em}I}$ is realized.
On the other hand, if $P_{\rm cg}$ is always less than unity, the system forms Onsager vortices.
Figure \ref{fig:cgp_rate} shows the time dependence of $P_{\rm cg}$, (a) and (b) show the results for $g_{12} \leqq 0.5g$ and $g_{12} \geqq 0.6g$, respectively.
In Fig. \ref{fig:cgp_rate} (a) $P_{\rm cg}$ is always less than unity from $t = 250$ to $1000$ for each $g_{12}$.
In Fig. \ref{fig:cgp_rate} (b), $P_{\rm cg}$ sometimes exceeds unity for each $g_{12}$.
Thus, we have Onsager vortices for $g_{12} < 0.5g$ and phase-separated vortices for $g_{12} > 0.6$.


\subsection{Dependence of the distribution of vortices on the interaction energies }\label{sec:energy}

This transition can be roughly explained by the dependence of the ratio of the energy of the intra-component and inter-component interaction of the vortices on $g_{12}$.
These two energies are calculated by Eq. (\ref{intracomponent_interaction_energy}) and (\ref{intercomponent_interaction_energy}).
Then, their ratio $P_{ij} = \epsilon^{\rm inter}_{ij}/\epsilon^{\rm intra}_{ij}$ becomes
\begin{equation}
P_{ij} = \frac{1}{4} \frac{g_{12}/g} {1 - (g_{12}/g)^{2}} \frac{\ln (\tilde {R'}_{ij}/2)}{(\tilde {R'}_{ij}/2)^{2}} \frac{1}{\ln (2R_{0}/\tilde {R}_{ij})}, \label{ratio_interaction}
\end{equation} 
with
\begin{equation}
\tilde{R}_{ij} = \frac{R_{ij}}{\xi} \ , \ \tilde{R'}_{ij} = \frac{2R'_{ij}}{\xi}. \label{normalize}
\end{equation}
The dependence of $P_{ij}$ on $g_{12}$ is represented by the factor 
\begin{equation}
\frac{g_{12}/g} {1 - (g_{12}/g)^{2}}, \nonumber
\end{equation} 
which becomes unity when $g_{12}/g \simeq 0.618$.
This means that interaction of the vortices which is dominant changes at approximately $g_{12}/g \simeq 0.618$.

The above arguments may be insufficient, because the interaction energy depends on the configuration of the vortices.
We should consider the sums $E^{\rm intra}$ and $E^{\rm inter}$ of the energies of the intra-component and inter-component interaction $\epsilon^{\rm intra}_{ij}$ and $\epsilon^{\rm inter}_{ij}$ of the vortices.
Let us denote the energy of the intra-component interaction of a pair of vortices of the $k$-th component as $\epsilon^{\rm intra}_{ij, k}$, and then $E^{\rm intra}$ and $E^{\rm inter}$ are calculated by 
\begin{equation}
E^{\rm intra} = \sum_{i<j}\epsilon^{\rm intra}_{ij, 1} + \sum_{i'<j'}\epsilon^{\rm intra}_{i'j', 2}, \label{total_intra}
\end{equation}
\begin{equation}
E^{\rm inter} = \sum_{i,j}\epsilon^{\rm inter}_{ij}, \label{total_inter}
\end{equation}
respectively.
Here, $\epsilon^{\rm inter}_{ij}$ is negative when $\tilde{R'}_{ij} < 2$ (Eqs. (\ref{intercomponent_interaction_energy}) and (\ref{normalize})).
Then the vortices overlap each other, because their distance is less than $2\xi$.
However, the contribution of this situation to the system is not so relevant when we consider phase-separated vortices, because the vortices of different components seldom overlap each other, except for the boundary of the VRs.
Thus, we sum up only the pairs with $\tilde{R'}_{ij} > 2$ in Eq. (\ref{total_inter}).

We can predict that the system forms Onsager vortices when $E^{\rm intra} \gg E^{\rm inter}$.
The transition to phase-separated vortices should occur at $E^{\rm intra} \simeq E^{\rm inter}$, which means that the inter-component interaction strongly affects the dynamics of vortices.

\begin{center}
\begin{figure}[h]
\includegraphics[width=20pc, height=10pc]{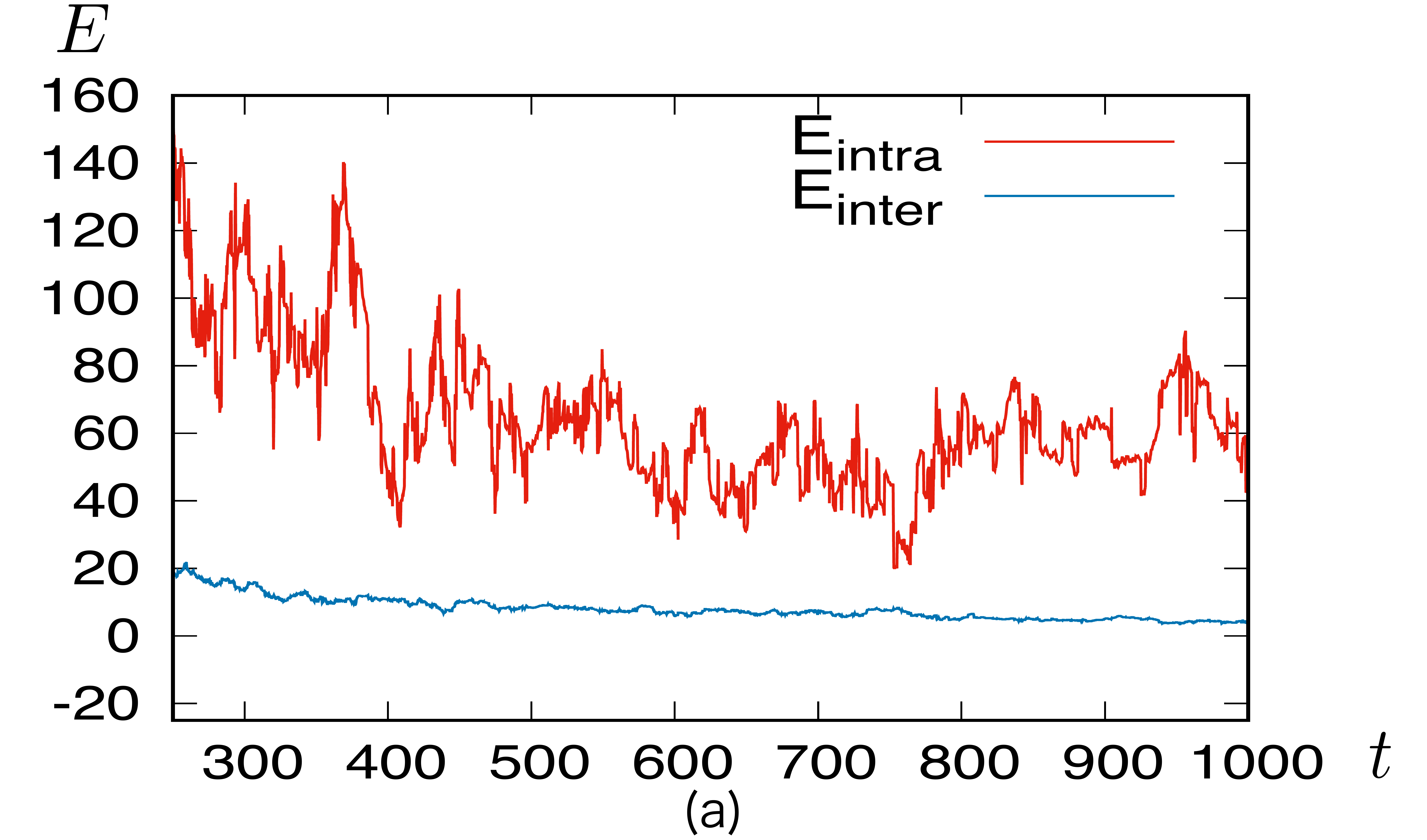}
\includegraphics[width=20pc, height=10pc]{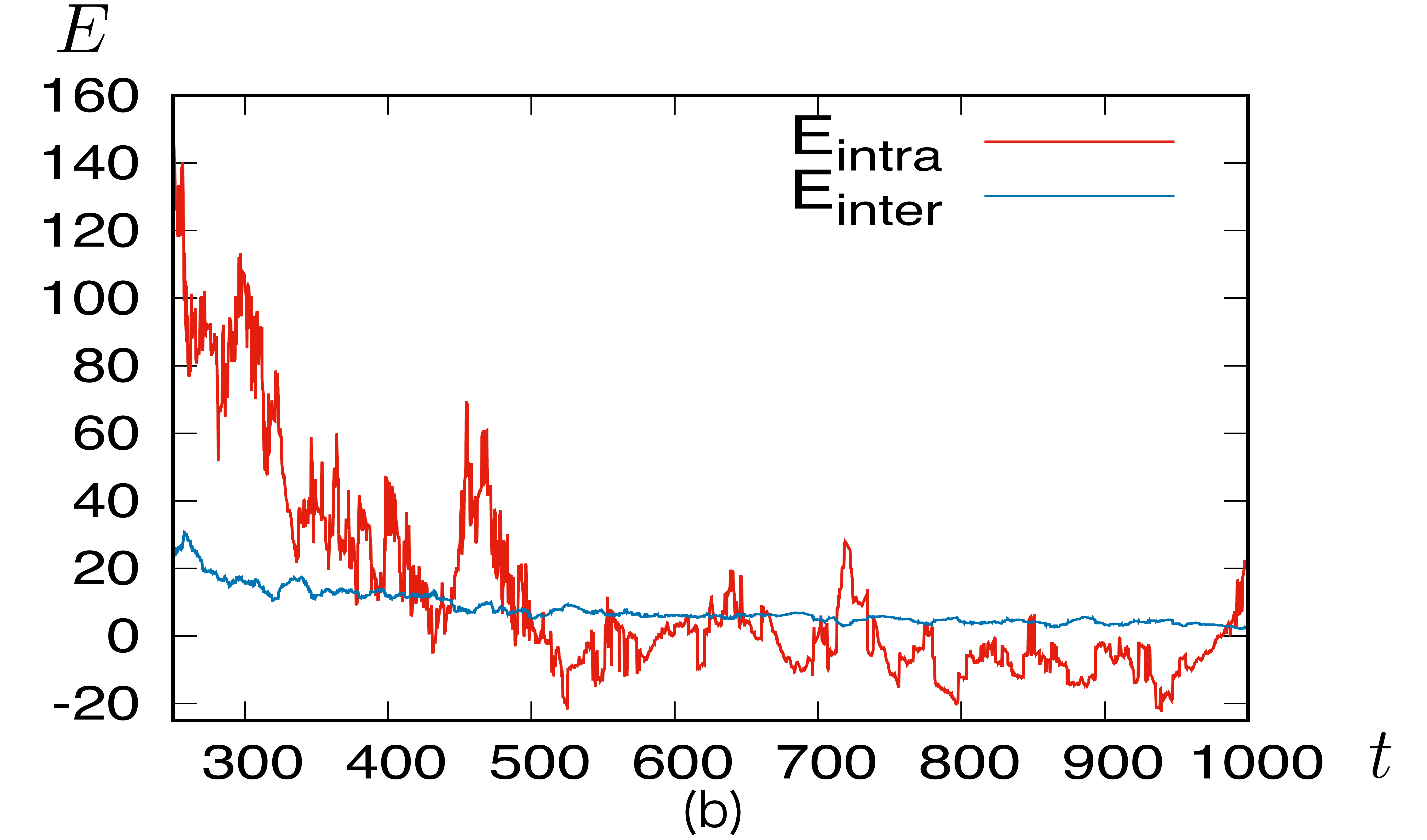}
\includegraphics[width=20pc, height=10pc]{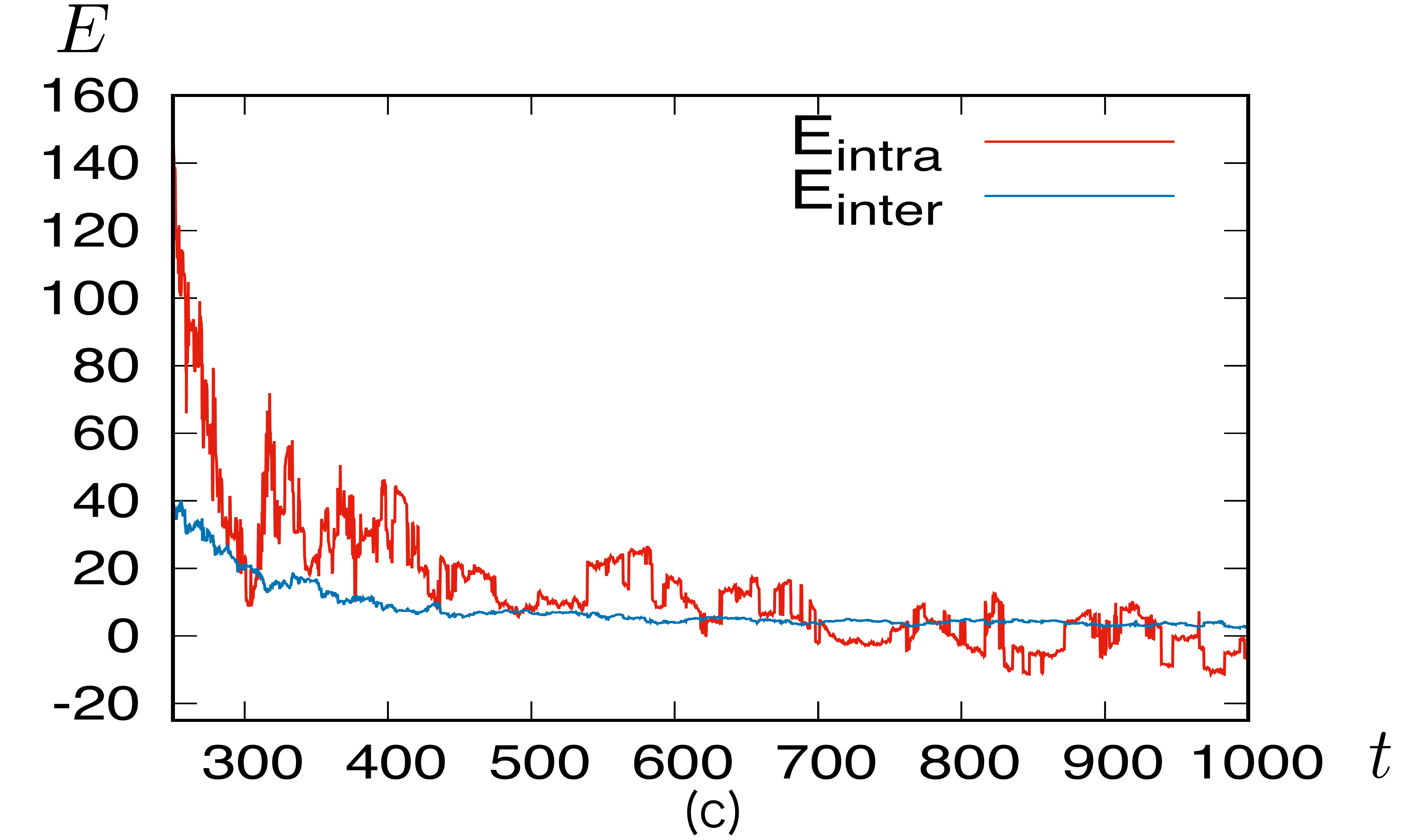}
\caption{\label{fig:int} Time dependence of the total energies $E^{\rm intra}$ and $E^{\rm inter}$ of the interaction of vortices with (a) $g_{12} = 0.5g$, (b) $g_{12} = 0.6g$, and (c) $g_{12} = 0.7g$.
The red and blue lines denote $E^{\rm intra}$ and $E^{\rm inter}$, respectively.}
\end{figure}
\end{center}

The time dependence of $E^{\rm intra}$ and $E^{\rm inter}$ is shown in Fig. \ref{fig:int} for three different values of $g_{12}$.
Here, $E^{\rm intra}$ fluctuate dramatically, while $E^{\rm inter}$ fluctuate within narrow limits.
This is derived from the dependence of Eqs. (\ref{intracomponent_interaction_energy}) and (\ref{intercomponent_interaction_energy}) on the change in the distances between vortices.
The differentiations of Eqs. (\ref{intracomponent_interaction_energy}) and (\ref{intercomponent_interaction_energy}) with respect to the distance are
\begin{equation}
\frac{\partial}{\partial \tilde{R}_{ij}} \epsilon^{\rm intra} \propto \frac{1}{\tilde{R}_{ij}}, \nonumber
\end{equation}
\begin{equation}
\frac{\partial}{\partial \tilde{R'}_{ij}} \epsilon^{\rm inter} \propto \frac{1}{(\tilde{R'}_{ij}/2)^{3}}\left(1-\ln(\tilde{R'}_{ij}/2)\right), \nonumber
\end{equation}
respectively.
Because the pairs $\tilde{R'}_{ij} < 2$ are cut off in Eq. (\ref{total_inter}), and the averages of $\tilde{R}_{ij}$ and $\tilde{R'}_{ij}$ are approximately $10$, $\epsilon^{\rm intra}$ is more sensitive to distance changes than $\epsilon^{\rm inter}$.
On the three panels, $E^{\rm intra}$ decreases to $t \simeq 400$ and then fluctuates around some values.
Here $t \simeq 400$ is the time when the decrease in the number of vortices caused by the annihilation of pairs almost stops (Figs. \ref{fig:2C_Nv_dipole_101} (a) and \ref{fig:2C_Nv_dipole_107} (a)).
Thus, the decrease in $E^{\rm intra}$ to $t \simeq 400$ occurs due to the decrease in the number of pairs of vortices, and the fluctuation after $t \simeq 400$ is caused by the time dependence of distribution of vortices.
When $g_{12}=0.5g$ (Fig. \ref{fig:int} (a)), $E^{\rm intra}$ is always greater than $E^{\rm inter}$.
When $g_{12} = 0.6g$ (Fig. \ref{fig:int} (b)) and $g_{12} = 0.7g$ (Fig. \ref{fig:int} (c)), $E^{\rm intra} \simeq E^{\rm inter}$ after $t \simeq 400$.
From these results, we can conclude that when $E^{\rm intra} \gg E^{\rm inter}$, the system has Onsager vortices, and when $E^{\rm intra} \simeq E^{\rm inter}$, the system has phase-separated vortices.
The transition from the Onsager vortices to phase-separated vortices occurs around $g_{12} \simeq 0.6g$, which is consistent with the prediction of this section, where the conclusion is derived from the time dependence of $P_{\rm cg}$ and the estimation of $P_{ij}$.


\section{Conclusions} \label{sec:Conclusions}

We have studied the dynamics of quantized vortices in two-dimensional two-component miscible BECs trapped by a box potential.
In this case, the dynamics of vortices strongly depends on the inter-component coupling $g_{12}$.
When $g_{12}$ is small, the formation of Onsager vortices is confirmed in both components, which is consistent with the case of one-component BECs.
When $g_{12}$ is large, the system shows the phase separation of the distribution of vortices.
This is a phenomenon in which the vortex regions occupied by the vortices of each component are spatially separated by the inter-component interaction of the vortices.
These two types of distributions of vortices can be distinguished by the time dependence of the effective size ratio of VR and the distance between the CGVs of each component.
The transition between these two types is predicted to occur around $g_{12} \simeq 0.618g$, which is understood from the dependence of the ratio of the energies of the intra-component and inter-component interaction of vortices on $g_{12}$. 
We also calculate the total energy of interaction of vortices.
Whether the total intra-component energy is much greater than the total inter-component energy or not decides whether the system has Onsager vortices or phase-separated vortices.

Formation of Onsager vortices means that the vortices construct large-scale structures in both one-component and two-components BECs.
On the other hand, the phase separation of the distribution of vortices suppresses large-scale structures and is peculiar to multi-component BECs.
Then they make the dynamics of vortices of multicomponent BECs more complex than the dynamics of one-component BECs, and it is important to understand these two types of phenomena to study multicomponent turbulence.
We should make comments on the possibility of observing these phenomena.
The initial state of the random distribution of vortices can be prepared using phase imprinting.
The sign of the circulation of vortices is confirmed using the Bragg scattering method \cite{Seo17}, and thus, Onsager vortices and phase-separated vortices can be observed directly.
The inter-component coupling can be changed by Feshbach resonance to confirm the transition between them; then these phenomena should be observed.


\begin{acknowledgments}
This work was supported by JSPS KAKENHI Grant No. 17K05548.
\end{acknowledgments}

\bibliography{book,aps,other}

\end{document}